\documentclass[11pt]{article}
 \usepackage{amssymb}
 \usepackage{amsmath}
 \usepackage{latexsym}
 \usepackage[cp866]{inputenc}
 \usepackage[english]{babel}
 \usepackage[dvips]{graphicx}
 \usepackage{float}

\begin{document}
\textwidth 17cm
\textheight 24cm
\topmargin -1.5cm

\title{\bfseries \textup{Scaling of gamma-spectra registered by 
semiconductor detectors}}
\date{\small {(Received \today)}}
\author{\bfseries \textit{E. G. Obrazovskii}\footnote{E-mail:
e{\_}obrazovskii@ngs.ru}  \\
\small \itshape  Novosibirsk State University,
\small \itshape  630090, Novosibirsk, Russia }

\maketitle

\begin{abstract}

The scaling properties of   gamma-spectra recorded by semiconductor 
detectors are investigated. For practical purposes the method of simulation
multicomponent spectra using single experimental spectrum are suggested.

\end{abstract}

\begin{flushleft}
Pacs number(s): 07.85.Nc, 29.90.Kv, 29.40.Wk, 82.80.Ej
\end{flushleft}
















\maketitle

High-resolution gamma-spectrometry is a powerful tool in different
fields of science and technology: from astrophysics \cite{bib:Astr} and nuclear
spectrometry \cite{bib:NucSpectr} to gamma-ray imaging 
spectrometers \cite{bib:Imagers}, environmental science \cite{bib:Environment} 
and neutron activation analysis \cite{bib:NAA}. The Ge-detectors are most common
used in gamma-spectrometry due its high energy resolution and sufficiently
high efficiency registration \cite{bib:Baldin}. In recent years significant progress 
has been made in the technology of semiconductor detectors made of materials
with high atomic numbers (CdTe, HgI$_2$) \cite{bib:CdTeDet}.

The spectrum   registered by semiconductor detectors (the detector response
to a gamma\d ray)  consist of not only full-energy peak, but
also continuum distribution resulting from Compton scattering of primary
gammas and subsequent escaping of the scattered gammas from 
detector \cite{bib:Baldin}. 
This continuum severely limits 
the accuracy of the determination of peak areas in low-energy part in 
multicomponent gamma spectra.
The main physical processes of energy dissipation in the semiconductor detectors
are well-known \cite{bib:Baldin} and are the basis for numerious methods 
for Monte Carlo simulation gamma-spectra \cite{bib:Simulation}.
To apply these methods require detailed  
information about the exact geometrical parameters of both the detector crystal 
and   surrounding material \cite{bib:Karamanis}.

In this paper we study the scaling properties of   gamma-spectra 
registered by semiconductor detectors. For practical purposes the 
method of simulation multi-component spectra with using single experimental 
spectrum is suggested.

The most simplest scaling properties have so-called planar detectors with
sensitive volume $\sim 1 cm^3$, since the main contribution to the formation
of a continuous distribution gives a single Compton scattering, while the 
multiple scattering contributes only a small correction.		
The single Compton scattering  dominates in region from $0$ to $E_g^{(1)} = E_0/(1+m c^2/2E_0)$,
where $E_0$ is the energy of primary gammas. The energy  of scattered gamma ray 
is determined by the  well-known equation \cite{bib:LLQED}
\begin{equation}
  \frac{1}{\omega'_1} - \frac{1}{\omega_{01}} = 1 - \cos \theta,
\end{equation}
where for simplicity  $\omega_{01} \equiv E_0/m c^2$ and 
$m c^2= 511.0$\,keV  is the energy rest of the electron,
$\theta$ is the angle of scattering.
We can connect the probability of Compton scattering and total attenuation 
of gammas for two different energies of primary gammas, $E_{01}$ and $E_{02}$, 
for the same scattering angles, which corresponds to the condition
\begin{equation}
  \frac{1}{\omega'_1} - \frac{1}{\omega_{01}} = 
  \frac{1}{\omega'_2} - \frac{1}{\omega_{02}}.
\end{equation}
Then the count in gamma spectra with energy $E_1=\omega_{01}-\omega'_1$,
deposited in detector for primary gammas with energy $E_{01}$, corresponds
to the count with energy $E_2=\omega_{02}-\omega'_2$,
deposited in detector for primary gammas with energy $E_{02}$, where
\begin{equation}
 E_2= \omega_{02} - \omega'_2=  
 \frac{E_1 \omega_{02}^2 }{\omega_{01} \omega_{02} + 
 (\omega_{01} -E_1)(\omega_{01}-\omega_{02})}.
\end{equation}

The ratio of the number of counts per unit energy interval of the two spectra 
(for the same activity of primary gammas) is determined 
as the ratio of differential cross-section for Compton 
scattering  \cite{bib:LLQED}
\begin{equation}
 \frac{d \sigma}{d \omega'}= \frac{\pi r_e^2}{\omega_0^2} 
 \left[\frac{\omega_{0}}{ \omega'}+\frac{\omega'}{ \omega_0} + 
 \left(\frac{1}{\omega'} - \frac{1}{\omega_{0}}\right)^2 - 
 2\left( \frac{1}{\omega'} - \frac{1}{\omega_{0}}\right) \right],
\end{equation}
where $r_e=e^2/mc^2$,
and the ratio of the total  attenuation coefficients of the incident and 
scattering gammas by detector material.
For angles of scattering   $\theta=0$ and $\theta=\pi$ these coefficients
are determined only by the height of the  detector $h$
\begin{align}
 K(\theta=0)&= e^{- \mu_{tot}(\omega_{0}) h }, \label{eq:5}\\
 K(\theta=\pi)&= \frac{1-
 e^{ - \left(\mu_{tot}(\omega_{0}) + \mu_{tot}(\omega') \right)h }}{h 
 \left(\mu_{tot}(\omega_{0}) + \mu_{tot}(\omega') \right) },\label{eq:6}
\end{align}
where  $\mu_{tot}(\omega)$ is the linear coefficient attenuation for
energy $\omega$, and $\omega_0$ is the energy of the incident gammas,
$\omega'=\omega_0/(1+2 \omega_0)$. 

\begin{figure}[t!]
 \includegraphics[width=0.75\textwidth,height=3in]{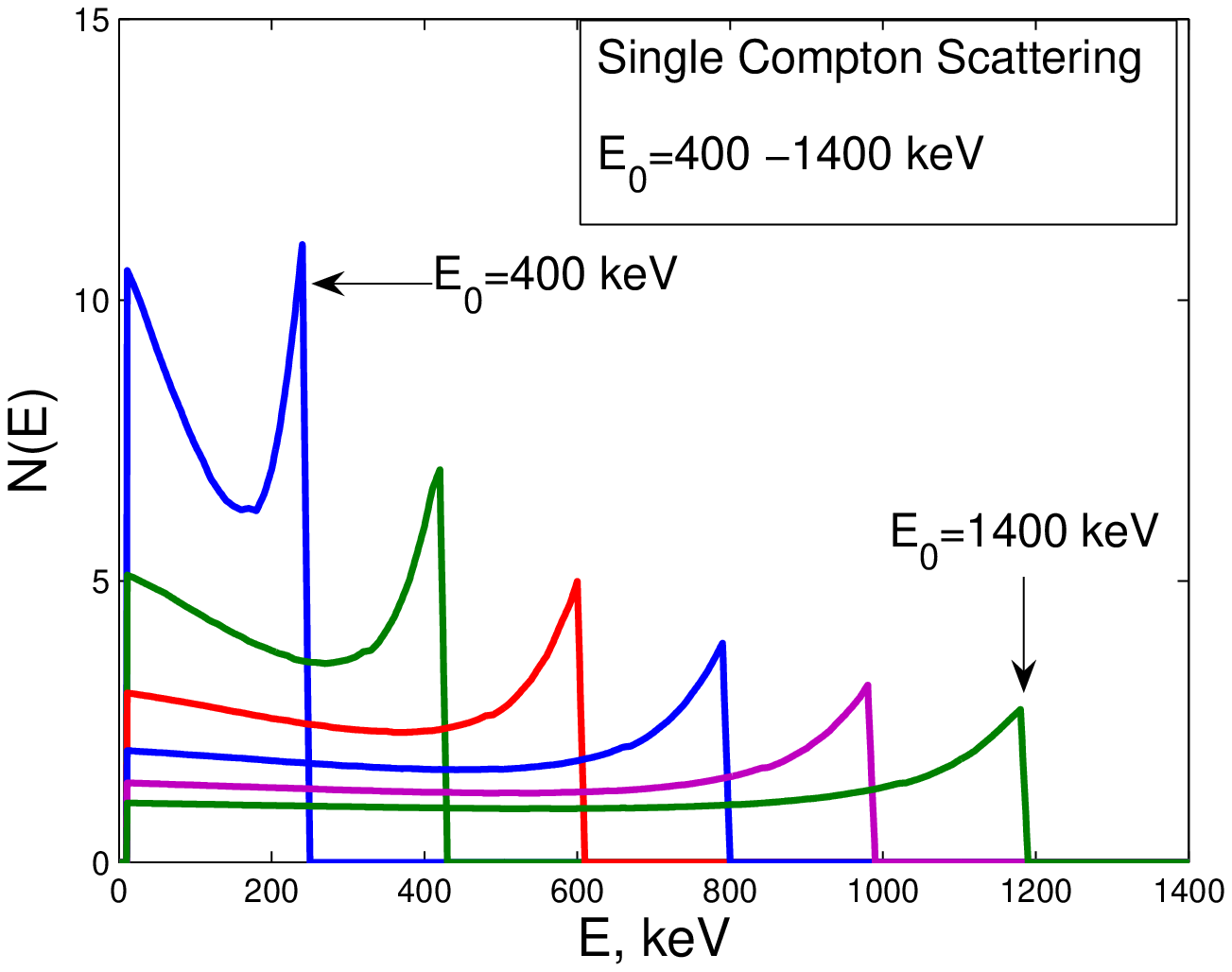}
 \hspace{2mm}
 \includegraphics[width=0.75\textwidth,height=3in]{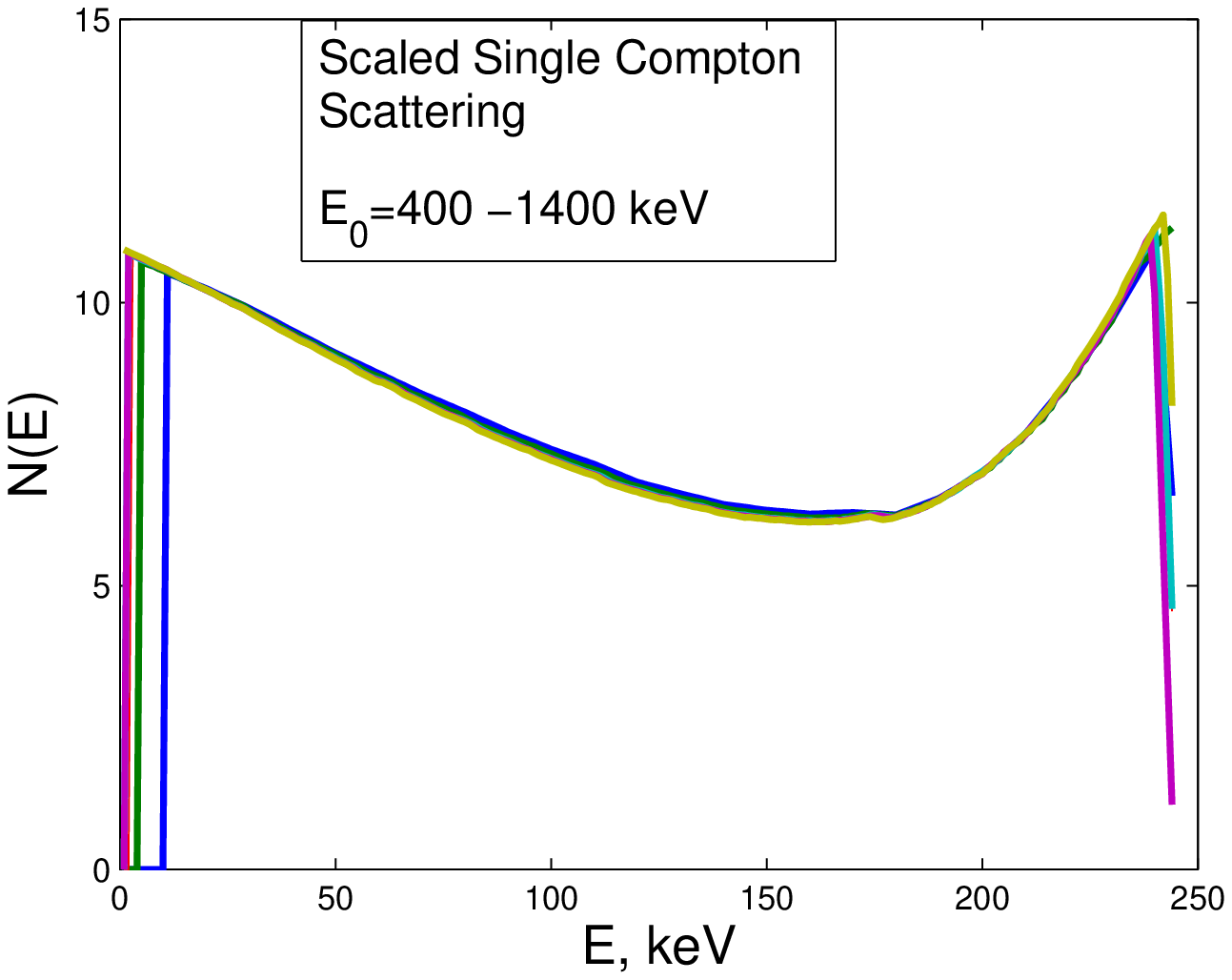}
 \caption{\small  Data numerical simulation of a single  Compton scattering
  spectra  for different energies of incident gammas ({\it top})
  and its scaling transformation ({\it bottom})}\label{SingleScatt}
\end{figure}

Fig.~\ref{SingleScatt} shows the data of numerical simulation of the spectra
of single Compton scattering of the planar Ge-detector ($\varnothing=1.0\,cm$, 
height $h=0.7\,cm$), and the scaling transformation of these spectra 
according to the
\begin{align}
 E_1 \to E_2&=\frac{E_1 \omega_{02}^2 }{\omega_{01} \omega_{02} + 
 (\omega_{01} -E_1)(\omega_{01}-\omega_{02})}, \\
 N_2(E_2) &= N_1(E_1)  \frac{d \sigma/ d \omega'_2}{d \sigma/ d \omega'_1}
 \frac{K(E_2)}{K(E_1)},
\end{align}
where the values of attenuation coefficients $K(E)$ for intermediate
energy values $E$ obtained by linear interpolation of the coefficients
(\ref{eq:5}) and (\ref{eq:6}).

\begin{figure}[t!]
 \includegraphics[width=0.75\textwidth,height=3in]{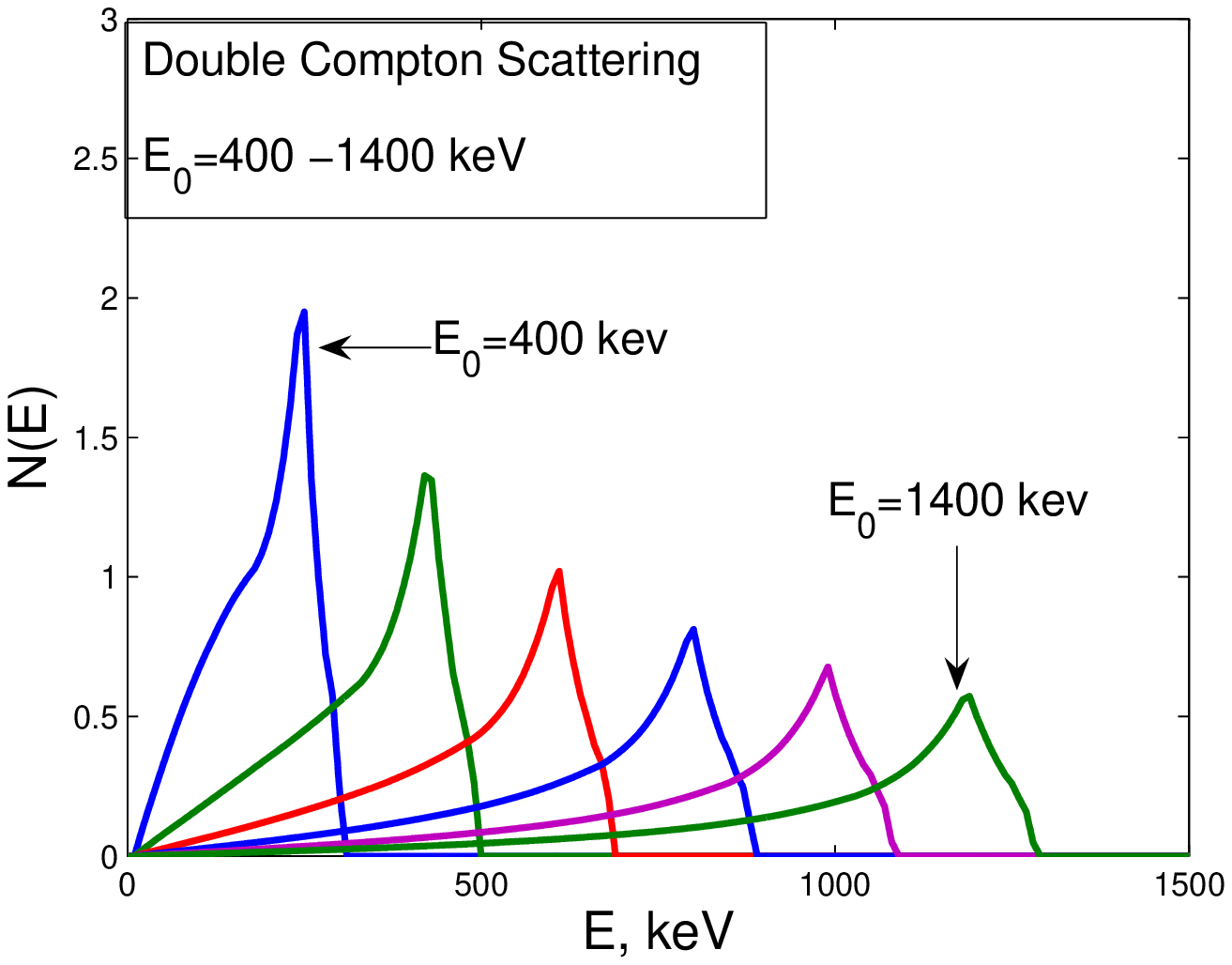}
 \hspace{2mm}
 \includegraphics[width=0.75\textwidth,height=3in]{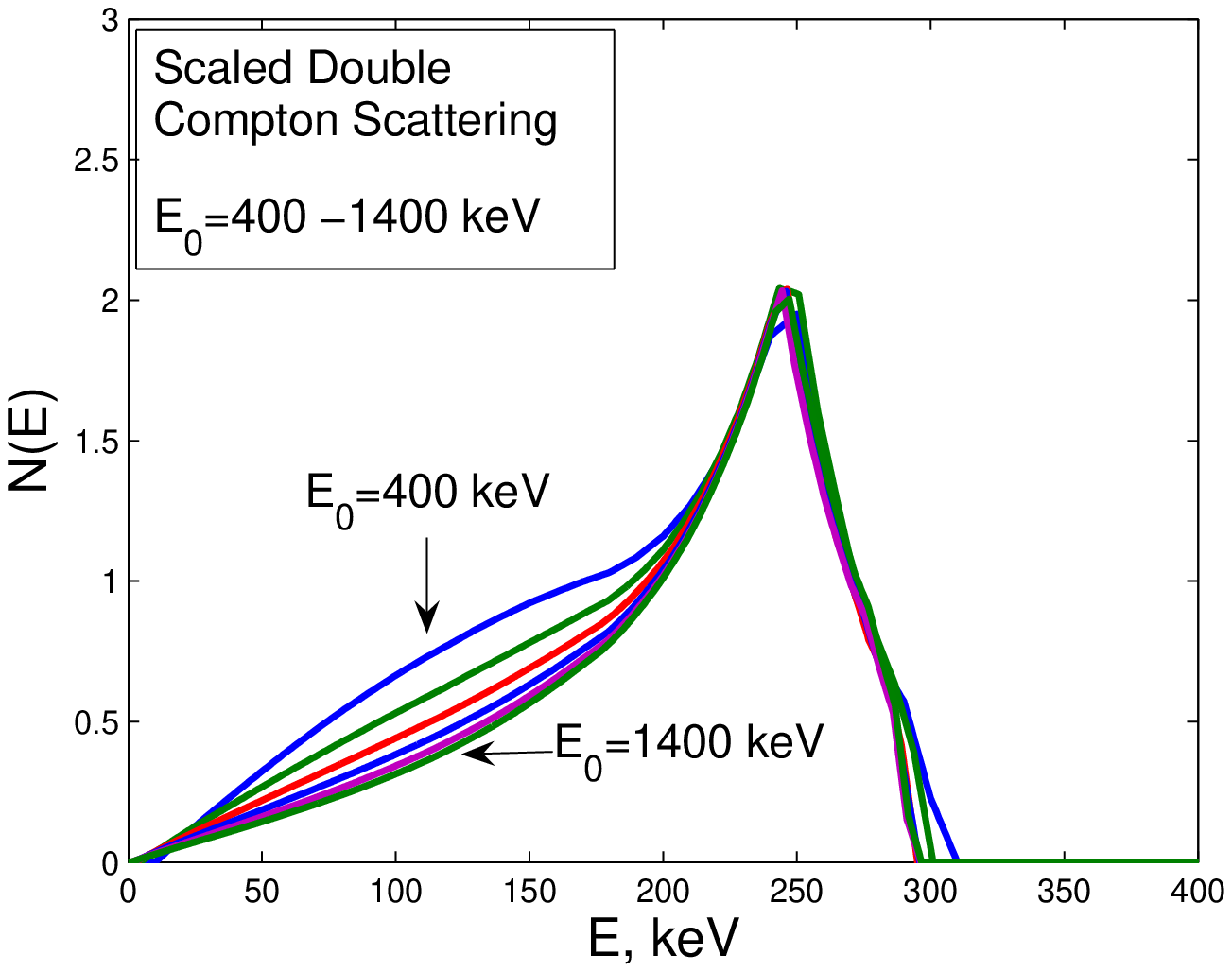}
 \caption{\small  Data numerical simulation of a double  Compton scattering
  spectra  for different energies of incident gammas ({\it top})
  and its scaling transformation ({\it bottom})}\label{DubbleScatt}
\end{figure}

In the region gamma-spectra from 
$E_g^{(1)}=\omega_0/(1+1/(2\omega_0))$ to 
$E_g^{(2)}=\omega_0/(1+1/(4\omega_0))$ the double Compton scattering 
is dominated. 
Continuum of  double scattering  has a maximum  at
$E_g^{(1)}$ and  approximately scales with the energy of the incident
gammas as $1/\omega_0$ that can be seen in fig.~\ref{DubbleScatt}, 
which shows the simulation data (top) and scaling  transformation (bottom).

The continuum from $E_g^{(2)}$ to $E_0$ associated with the scattering 
of the incident gamma radiation on the material of the enclosure and the 
detector entrance window at a small angles with complete absorption of the 
scattered gamma-ray photon by the detector.

Compton-scattered gammas in the material surrounding detector at
angles $\theta \to \pi$, then completely absorbed by the detector,
lead to backscatter peak. This peak is
scaled approximately as
\begin{align}
 E_{1b} \to E_{2b} &=  
 \frac{E_{1b} \omega_{01} \omega_{02}}{\omega_{01} \omega_{02} + 
 E_{1b}(\omega_{01}-\omega_{02})},  \\ 
 N_2(E_{2b}) &\to N_1(E_{1b})  \frac{d \sigma/ d \omega'_2}{d \sigma/ d \omega'_1}.
\end{align}

For many applications, the electron-positron pair production gives 
a negligible contribution to the Compton background Ge-detector for
energies of  gammas less than $3-5$\,MeV \cite{bib:Baldin}.

For practical application of established scaling laws the 
method for simulation multicomponent spectra, using the experimental
spectrum of a monoenergetic source and the detection efficiency of
total absorption peaks, are suggested.

\begin{figure}[t!]
 \includegraphics[width=0.75\textwidth,height=3in]{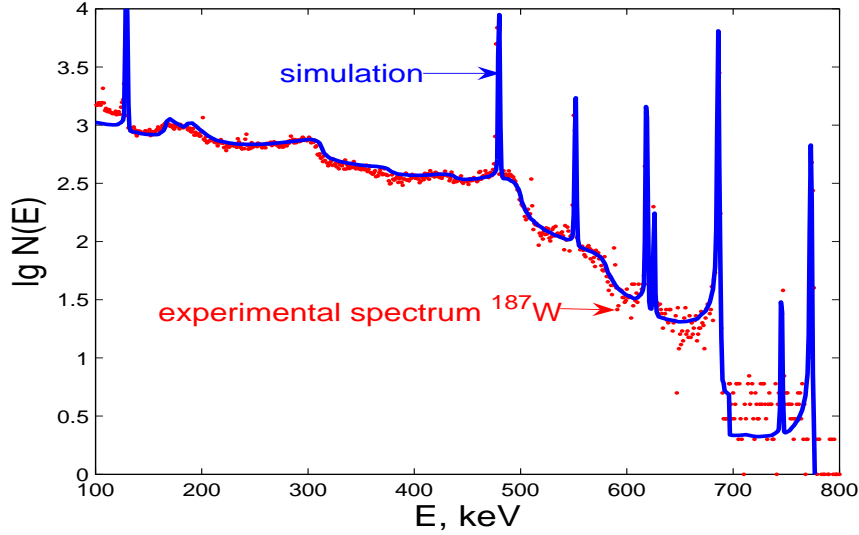}
 \caption{\small Comparison of experimental ({\it dots}) and simulation 
 ({\it line}) spectra  ${}^{187}$W.}\label{SimulationSpectra}
\end{figure}

As an example, the  fig.~\ref{SimulationSpectra} shows a comparison 
of the experimental spectrum of the radionuclide ${}^{187}$W,
measured using Ge-detector (ORTEC, 1013-10190, $\varnothing=1.0\,cm$, 
height $h=0.7\,cm$) and its simulation with spectrum of the radionuclide
${}^{198}$Au.
For this purpose, from the spectrum of the radionuclide ${}^{198}$Au was removed as the 
backscattered peak (by subtracting Compton background), as part of
double Compton scattering, which has energy higher than $E_g^{(1)}$.
These components of the spectrum were transformed using established 
for each of them above scaling laws. The peak of total absorption was converted 
in accordance with the detection efficiency, as measured with the use
of the radionuclide ${}^{182}$Ta. Then all the contributions were summed 
for each gamma-line of the simulated spectrum.

In conclusion, the scaling properties of   gamma-spectra registered by 
semiconductor detectors are investigated. A method for simulation 
multi-component spectra using scaling transformation of the  experimental
spectrum of a monoenergetic radiation is suggested.
Technical details of modeling the spectra and discussion of the 
approximations made will be the subject of a separate publication.

This work was partially carried out within the framework  Development
Programm of the National Research University Novosibirsk State University
at 2009 -- 2018 years.

\end{document}